\documentclass[epj,nopacs]{svjour}

\usepackage[T1]{fontenc}
\usepackage[affil-it]{authblk}
\usepackage{amsmath,amssymb,mathrsfs,mathtools,array,booktabs}
\usepackage{newtxtext,newtxmath,wrapfig,slashed,soul,enumitem}
\usepackage[dvipsnames]{xcolor}
\usepackage[caption=false,singlelinecheck=false]{subfig}
\usepackage{marginnote,setspace,cite,placeins,soul}
\usepackage[colorlinks, urlcolor=black, citecolor=blue, link color=black]{hyperref}

\renewcommand{\d}{\ensuremath{\mathrm{d}}}
\newcommand{\modulus}[1]{\ensuremath{\big| #1 \big|}}
\newcommand{\alphaem}{\ensuremath{\alpha_{\textrm{EM}}}}
\newcommand{\SM}{\ensuremath{\textrm{SM}}}

\newcommand{\Br}{\ensuremath{\textrm{Br}}}
\newcommand{\Kbar}{\ensuremath{\overline{K}}}
\newcommand{\Bbar}{\ensuremath{\overline{B}}}

\numberwithin{equation}{section}
\journalname{}

\def\makeheadbox{ }

\allowdisplaybreaks

\begin{document}

\title{\bf Searching for signatures of new physics in \texorpdfstring{$\boldsymbol{B \to K \, \nu \, \overline{\nu}}$}{B -> K nu nubar} to distinguish between Dirac and Majorana neutrinos}

\titlerunning{Signatures of NP in $B \to K \, \nu \, \overline{\nu}$ to distinguish Dirac and Majorana neutrinos}

\author{C.~S.~Kim\inst{1}\thanks{cskim@yonsei.ac.kr}, Dibyakrupa Sahoo\inst{2}\thanks{Dibyakrupa.Sahoo@fuw.edu.pl (corresponding author)} \and K.~N.~Vishnudath\inst{3}\thanks{vishnudath.neelakand@usm.cl}}

\authorrunning{Kim, Sahoo and Vishnudath}

\institute{Department of Physics and IPAP, Yonsei University, Seoul
03722, Korea \and Institute of Theoretical Physics, Faculty of
Physics, University of Warsaw, ul.\ Pasteura 5, 02-093 Warsaw, Poland
\and Universidad T\'{e}cnica Federico Santa Mar\'{i}a, Casilla 110-V,
Valpara\'{i}so, Chile}

\date{\today}%

\abstract{We conduct a model-independent analysis of the distinct
signatures of various generic new physics possibilities in the decay
$B \to K \, \nu \, \overline{\nu}$ by analyzing the branching ratio as
well as the missing mass-square distribution. Considering the final
neutrinos to be of the same flavor with non-zero mass, we discuss the
new physics contributions for both Dirac and Majorana neutrino
possibilities. In our study, we utilize the analytical relations among
form factors in semi-leptonic $B \to K$ transitions, which are
consistent with current lattice QCD predictions to a very high
numerical accuracy. We provide constraints on different new physics
parameters, taking into account the recent measurement of $B^+ \to K^+
\, \nu \, \overline{\nu}$ branching ratio by the Belle-II
collaboration. In future, if the missing mass-square distribution for
$B^+ \to K^+ \, \nu \, \overline{\nu}$ decay gets reported by Belle-II
with analysis of more events than their present data set, one can not
only investigate possible new physics effects in these decays, but
also probe the Dirac/Majorana nature of the neutrinos using quantum
statistics, since a difference between the two cases is known to exist
in the presence of non-standard neutrino interactions.}

\maketitle

\section{Introduction}\label{sec:intro}

The flavor changing neutral current (FCNC) processes involving the
quark-level transition $b \to s$ are facilitated in the standard model
(SM) only via quantum loop effects which are suppressed by the
Glashow-Iliopoulos-Maiani mechanism \cite{Glashow:1970gm}. Theoretical
and experimental studies in this regard have thoroughly explored
various avenues, such as the rare mesonic decays $B \to K^{(*)} \,
\ell^+ \, \ell^-$, $B \to K^{(*)} \, \nu_\ell\, \overline{\nu}_\ell$,
$B_s \to \mu^+ \, \mu^-$, $B_s \to \phi \, \ell^+ \, \ell^-$ with
$\ell=e,\mu,\tau$ (for example, see \cite{Henderson:2023occ} and
\cite{Bifani:2018zmi} along with the references therein) as well as
the baryonic decay $\Lambda_b \to \Lambda \, \mu^+ \, \mu^-$
\cite{CDF:2011buy, LHCb:2015tgy}. These studies are focused on finding
any deviations from the SM predictions as a way to study different new
physics (NP) possibilities, which can contribute either in loops or
even at the tree-level \cite{Kim:1999waa, Kim:2018hlp}. In
model-independent analyses these NP contributions are usually
parameterized in terms of various effective operators. Experimentally,
mesonic decays involving pair of charged leptons ($\ell^+ \, \ell^-$)
in the final state are easier to study. However, these decays also
involve contributions due to photon exchange, a feature which breaks
factorization and leads to significant hadronic uncertainties beyond
form factors. In addition, the presence of intermediate hadronic
resonances such as $J/\psi$, $\psi(2S),$ etc.,\ also affect the study
of such processes \cite{Khodjamirian:2010vf}. However, in the case of
$B \to K \, \nu \, \overline{\nu}$, the only hadronic uncertainties
come from form factors with no extra intermediate resonance effects.
The $B \to K$ form factors are obtained from studies involving various
non-perturbative methods such as lattice QCD and light-cone sum rules
(LCSR). In the SM the branching ratio of the decay $B^+ \to K^+ \, \nu
\, \overline{\nu}$ is predicted to be \cite{Belle-II:2023esi},
\begin{equation}\label{eq:BrSM}
\Br\left(B^+ \to K^+ \, \nu \, \overline{\nu}\right)_{\SM} =
\left(4.97 \pm 0.37\right) \times 10^{-6},
\end{equation}
from loop contributions alone. For this decay there is a possible
tree-level background contribution in the SM arising from the
sequential decay $B^+ \to \tau^+ \left(\to K^+ \,
\overline{\nu}_\tau\right) \, \nu_\tau$ \cite{Parrott:2022zte} which
involves doubly-charged-current interaction. According to
\cite{Belle-II:2023esi}, the branching ratio for
this long-distance contribution is,
\begin{equation}\label{eq:BrSMTreeBackground}
\Br\left(B^+ \to \tau^+  \, \nu_\tau \to K^+  \, \nu_\tau \,
\overline{\nu}_\tau\right)_{\SM} = \left(0.61 \pm 0.06\right) \times
10^{-6},
\end{equation}
which is roughly ten percent of the SM loop-induced branching ratio.
Despite the apparent theoretical advantages in $B^+ \to K^+ \, \nu \,
\overline{\nu}$, the presence of a pair of invisible (unreconstructed)
neutrinos in the final state offers a formidable experimental
challenge to observe this process and distinguish it from various
possible background processes.

Notwithstanding the various experimental difficulties, the Belle-II
collaboration has recently reported \cite{Belle-II:2023esi} the first
evidence for experimental observation of $B^+ \to K^+ \, \nu \,
\overline{\nu}$ and reported its measured branching ratio to be
\begin{equation}\label{eq:BrBelleII}
\Br(B^+ \to K^+ \, \nu \, \overline{\nu})_{\textrm{Belle-II}} =
\left(2.3 \pm 0.7\right) \times 10^{-5},
\end{equation}
which is $2.57\sigma$ bigger than the SM expectation from FCNC. The SM
allowed tree-level contribution is not large enough to account for
this deviation. However, considering the low significance of the
reported discrepancy, one can at best consider this as a tentative
hint for existence of different NP possibilities \cite{Bause:2023mfe,
Allwicher:2023xba, Athron:2023hmz, Felkl:2023ayn, He:2023bnk,
Chen:2023wpb, Datta:2023iln, Altmannshofer:2023hkn, McKeen:2023uzo,
Fridell:2023ssf, Berezhnoy:2023rxx, Ho:2024cwk, Chen:2024jlj,
Loparco:2024olo, Gabrielli:2024wys, Hou:2024vyw, Chen:2024cll,
He:2024iju, Bolton:2024egx, Marzocca:2024hua, Rosauro-Alcaraz:2024mvx,
Karmakar:2024gla, Buras:2024ewl}. Nevertheless, we are interested in
estimating the contributions from all generic NP possibilities allowed
by the current measurement. We conduct a model-independent analysis,
starting from the effective Lagrangian, considering  vector, scalar,
and tensor NP interactions. Moreover, we also study how the new
physics contributions vary based on whether the final state neutrinos
are Dirac or Majorana. Drawing from the analyses of \cite{Kim:2021dyj,
Kim:2022xjg, Kim:2023iwz}, we rely on a fundamental aspect of quantum
statistics: if the two final state neutrinos are Majorana particles,
they are indistinguishable from each other. Consequently, the decay
amplitude for Majorana case must be anti-symmetric with respect to the
exchange of the two neutrinos, resulting in a different decay width
compared to the Dirac case, in presence of NP. Taking one type of NP
interaction at a time (scalar, vector or tensor), we put constraints
on the relevant NP parameters for both Majorana and Dirac neutrino
possibilities, from the experimentally observed value of the $\Br(B^+
\to K^+ \, \nu \, \overline{\nu})$.

Before we proceed to our NP analysis in the decay $B \to K \nu
\overline{\nu}$, we would like to emphasize that our proposed method
of using quantum statistics is quite different from proposals that
search for signatures of lepton number violation (LNV). A summary
review of signatures and tests for Majorana nature of neutrinos using
LNV in rare $B$ and $K$ decays can be found in
Ref.~\cite{Buras:2023qaf}. It should be noted that Majorana neutrinos
can also lead to lepton number conservation just like Dirac neutrinos,
and the LNV signals are usually suppressed by powers of neutrino mass.
This makes our suggestion of utilizing quantum statistics a useful
alternative to the traditional LNV studies. 

We would also like to clarify that in principle one might
consider neutrino and antineutrino of different flavors in the final
state (i.e.\ consider $B \to K \nu_\ell \overline{\nu}_{\ell'}$ with
$\ell \neq \ell'$). Such lepton flavor violating decays are
experimentally impossible to distinguish and isolate from decays
having final neutrino and antineutrino of the same flavor (i.e.\ $B
\to K \nu_\ell \overline{\nu}_{\ell}$), since in both cases the
neutrino and antineutrino remain undetected in the detector
surrounding the place where the decays take place. We assume that the
experimentally observed decays are as expected in the SM. In the SM
allowed weak interaction lepton flavor is
conserved\footnote{Since we are considering the decay at its
point of production, the lepton flavor violation usually considered in
neutrino oscillations is not applicable in our case.} and is
universal. Therefore, in our model-independent analysis we focus on
lepton flavor conserving $B \to K \nu_\ell \overline{\nu}_{\ell}$
decays, and explore any possible NP effects in this specific context.
Considering lepton flavor conservation and lepton flavor universality
helps us to write down a minimal set of operators that essentially
leads to all possible variations in experimental observables which
would be of interest in future experimental studies of the $B \to K
\nu \overline{\nu}$ decays. Finally we note that sticking to lepton
flavor conservation, i.e.\ working with a neutrino-antineutrino pair
of the same flavor in the final state, is required to implement
antisymmetrization of the final state for Majorana nature of neutrino
when needed. In future when more experimental data get accessible, one
can extend our current approach by including lepton flavor violation
and lepton flavor non-universality as well.

Our paper is organized as follows. In Sec.~\ref{sec:NP-operators} we
introduce the set of generic new physics operators which can
contribute to $B \to K \, \nu \, \overline{\nu}$. Then in
Sec.~\ref{sec:form-factors} we provide a set of very simple analytical
expressions which relate the various form factors in $B \to K$
transition. Certainly the $B \to K$ form factors do play an important
role here. In our study, we use the analytical relations among the
form factors which fit the $B \to K$ lattice QCD results, as mentioned
in \cite{KPRS:2024}. Using these analytical relations, we study the
signatures of new physics in Sec.~\ref{sec:NP-signatures} more
efficiently. Finally we conclude in Sec.~\ref{sec:conclusion}
highlighting all the important findings of our study.

\section{Generic new physics possibilities in effective Lagrangian for \texorpdfstring{$\boldsymbol{B \to K \, \nu \, \overline{\nu}}$}{B -> K nu nubar}}\label{sec:NP-operators}

Considering all the new physics possibilities, the full effective
Lagrangian facilitating the decay $B \to K \, \nu_\ell \,
\overline{\nu}_\ell$ is given by,
\begin{equation}\label{eq:effective-Lagrangian-1}
\mathscr{L}_{\textrm{NP+SM}}^{b \to s \nu \overline{\nu}} = \frac{4 \,
G_F}{\sqrt{2}} \, \frac{\alphaem}{4\pi} \, \lambda_t \,
\sum_{\substack{{\ell=e,\mu,\tau}\\{Y=S,V,T}\\{X=L,R}}} \,
\left(C_X^Y\right)^{\ell} \, \left(\mathcal{O}_X^Y\right)^{\ell} +
\textrm{h.c.},
\end{equation}
where $G_F$ is the Fermi constant, $\alphaem$ is the electromagnetic
fine-structure constant, $\alphaem = e^2/(4\pi)$ with $e$ being the
electric charge of electron,
\begin{equation}
\lambda_t = V_{tb} \, V_{ts}^* = -0.0398 \pm 0.0008,
\end{equation}
with $V$ being the CKM matrix, $\left(C_X^Y\right)^{\ell}$ denote the
Wilson coefficients and the corresponding operators
$\left(\mathcal{O}_X^Y\right)^{\ell}$ are given by,
\begin{subequations}\label{eq:operators}
\begin{align}
\left( \mathcal{O}_L^V \right)^\ell &= \left(\overline{s} \,
\gamma^\mu \, b \right) \, \left( \overline{\nu}_\ell \, \gamma_\mu \,
P_L \, \nu_\ell\right),\\%
\left( \mathcal{O}_R^V \right)^\ell &= \left(\overline{s} \,
\gamma^\mu \, b \right) \, \left( \overline{\nu}_\ell \, \gamma_\mu \,
P_R \, \nu_\ell\right),\\%
\left( \mathcal{O}_L^S \right)^\ell &= \left(\overline{s} \, b \right)
\, \left( \overline{\nu}_\ell \, P_L \, \nu_\ell\right),\\%
\left( \mathcal{O}_R^S \right)^\ell &= \left(\overline{s} \, b \right)
\, \left( \overline{\nu}_\ell \, P_R \, \nu_\ell\right),\\%
\left( \mathcal{O}_L^T \right)^\ell &= \left(\overline{s} \,
\sigma^{\mu\nu} \, b \right) \, \left( \overline{\nu}_\ell \,
\sigma_{\mu\nu} \, P_L \, \nu_\ell\right),\\%
\left( \mathcal{O}_R^T \right)^\ell &= \left(\overline{s} \,
\sigma^{\mu\nu} \, b \right) \, \left( \overline{\nu}_\ell \,
\sigma_{\mu\nu} \, P_R \, \nu_\ell\right),%
\end{align}
\end{subequations}
with $P_L = \frac{1}{2} \, \left( 1-\gamma^5 \right)$ and $P_R =
\frac{1}{2} \, \left( 1+\gamma^5 \right)$ being the left-chiral and
right-chiral projection operators, respectively. We have not imposed
any constraints for Majorana nature of neutrinos in
Eq.~\eqref{eq:effective-Lagrangian-1}. We will do so later when we
compare the effects from Dirac and Majorana nature of neutrinos. It
should be noted that in the SM, we have \textit{lepton flavor
universality} and
\begin{equation}\label{eq:SM-Wilson-coeff}
\begin{aligned}
\left(C_L^V\right)^e_{\SM} &= \left(C_L^V\right)^\mu_{\SM} =
\left(C_L^V\right)^\tau_{\SM} = \dfrac{-X_t}{\sin^2\theta_W},\\%
\left(C_R^V\right)^\ell_{\SM} &= 0, \left(C_X^S\right)^\ell_{\SM} =
\left(C_X^T\right)^\ell_{\SM} = 0,
\end{aligned}
\end{equation}
for $X=L,R$ and $X_t = 1.469 \pm 0.017 \pm 0.002$ which includes the
NLO QCD corrections \cite{Buchalla:1993bv, Buchalla:1998ba,
Misiak:1999yg} as well as 2-loop electroweak contributions
\cite{Brod:2010hi}. Later when we consider NP effects, we assume that
the Wilson coefficients $\left(C_X^Y\right)^\ell$ deviate from the SM
values by a non-zero but small amount. For simplicity, in this work we
assume that NP also respects \textit{lepton flavor universality}.
Then, the Wilson coefficients can be simply denoted without the
superscript $\ell$, i.e.\
\begin{equation}
\left(C_X^Y\right)^e = \left(C_X^Y\right)^\mu =
\left(C_X^Y\right)^\tau \equiv C_X^Y.
\end{equation}
Therefore, the lepton or neutrino flavor universal effective
Lagrangian is given by,
\begin{equation}\label{eq:effective-Lagrangian-2}
\mathscr{L}_{\textrm{NP+SM}}^{b \to s \nu \overline{\nu}} = \frac{4 \,
G_F}{\sqrt{2}} \, \frac{\alphaem}{4\pi} \, \lambda_t \,
\sum_{\substack{{\ell=e,\mu,\tau}\\{Y=S,V,T}\\{X=L,R}}} \, C_X^Y \,
\left(\mathcal{O}_X^Y\right)^{\ell} + \textrm{h.c.}\, .
\end{equation}
Note that in writing the operators
$\left(\mathcal{O}_X^Y\right)^{\ell}$ as expressed in
Eq.~\eqref{eq:operators}, we have already taken into consideration the
fact that the quark-level currents should preserve parity since both
$B$ and $K$ are pseudo-scalar mesons. In this work, we do not consider
any specific NP model in particular (e.g.\ see
Refs.~\cite{Bause:2023mfe,Allwicher:2023xba, Athron:2023hmz,
Felkl:2023ayn, He:2023bnk, Chen:2023wpb, Datta:2023iln,
Altmannshofer:2023hkn, McKeen:2023uzo, Fridell:2023ssf,
Browder:2021hbl, Berezhnoy:2023rxx, Ho:2024cwk, Chen:2024jlj,
Loparco:2024olo, Gabrielli:2024wys, Hou:2024vyw, Chen:2024cll,
He:2024iju, Bolton:2024egx, Marzocca:2024hua, Rosauro-Alcaraz:2024mvx,
Buras:2024ewl} and references therein for specific NP model
considerations). While going from the quark-level effective Lagrangian
of Eq.~\eqref{eq:effective-Lagrangian-2} to the decay amplitude at the
meson level for $B \to K \, \nu_\ell \, \overline{\nu}_\ell$, one
usually encounters three different form factors which are evaluated by
taking non-perturbative QCD effects into consideration. These form
factors are functions of the momentum transfer-squared which is the
only available dynamic scalar quantity in this case. We introduce the
three form factors as well as analytical relations among them in the
following section.

\section{The \texorpdfstring{$\boldsymbol{B \to K}$}{B -> K} form factors and relations among them}\label{sec:form-factors}

The three dimensionless and real form factors one encounters in the $B
\to K$ semi-leptonic decays are the scalar form factor $f_0 (q^2)$,
the vector form factor $f_+ (q^2)$ and the tensor form factor $f_T
(q^2)$, which are defined by the following hadronic matrix elements,
\begin{subequations}\label{eq:hadronic-matrix-elements}
\begin{align}
&\big\langle \Kbar(p_K) \big\lvert \overline{s} \, b \big\rvert
\Bbar(p_B) \big\rangle = \left(\dfrac{m_B^2 - m_K^2}{m_b - m_s}\right)
\, f_0 (q^2), \label{eq:scalar-matrix-element}\\%
&\big\langle \Kbar(p_K) \big\lvert \overline{s} \, \gamma^\mu \, b
\big\rvert \Bbar(p_B) \big\rangle = p^\mu \, f_+ (q^2) \nonumber\\%
&\qquad - \left(\frac{m_B^2 - m_K^2}{q^2}\right) \, q^\mu \, \left(
f_+ (q^2) - f_0 (q^2) \right), \label{eq:vector-matrix-element}\\%
&\big\langle \Kbar(p_K) \big\lvert \overline{s} \, \sigma^{\mu\nu} \,
b \big\rvert \Bbar(p_B) \big\rangle = i \, \left(\frac{p^\mu \, q^\nu
- q^\mu \, p^\nu}{m_B + m_K}\right) \, f_T(q^2),
\label{eq:tensor-matrix-element}
\end{align}
\end{subequations}
where $p_B$, $p_K$ denote the 4-momenta of $B$ and $K$ mesons
respectively, with $p = p_B+ p_K$ and $q= p_B - p_K$ being the two
independent 4-momenta combinations,  $m_B$, $m_K$ denote the masses of
$B$ and $K$ mesons, and $m_b$, $m_s$ denote the masses of $b$ and $s$
quarks respectively.

\begin{figure}[hbtp]
\centering%
\includegraphics[width=\linewidth,keepaspectratio]{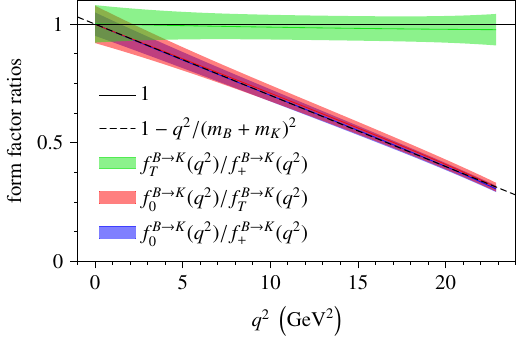}%
\caption{The form factor relations of Eqs.~\eqref{eq:f0fp-relation}
and \eqref{eq:fpfT-relation} give excellent numerical fit to the
lattice QCD form factors as predicted by HPQCD~\cite{Parrott:2022zte}
for $B \to K$ transition. The colored bands correspond to $1\sigma$
errors on the form factor ratios and are estimated by quadrature from
HPQCD form factor predictions.}%
\label{fig:FFs-LQCD}
\end{figure}

In the literature there exist various theoretical approaches such as
constituent quark model, heavy-quark effective theory, large energy
effective theory, light cone sum rules to analyze the form factors and
relations among them. The $B \to K$ form factors as predicted from
lattice QCD by the HPQCD collaboration \cite{Parrott:2022zte}, have
excellent numerical agreement (see Fig.~\ref{fig:FFs-LQCD}) with the
following form factors relations,
\begin{equation}\label{eq:f0fp-relation}
f_0(q^2) \approx \left(1-\frac{q^2}{(m_B + m_K)^2}\right) \, f_+(q^2),
\end{equation}
and
\begin{equation}\label{eq:fpfT-relation}
f_T(q^2) \approx f_+(q^2) .
\end{equation}
Such form factor relations can be found in the literature, e.g.\ in
Refs.~\cite{Stech:1995ec} and \cite{Soares:1996vs}. However, a more
detailed discussion and comparison of these and other existing form
factor relations are given in Ref.~\cite{KPRS:2024}. With
Eqs.~\eqref{eq:f0fp-relation} and \eqref{eq:fpfT-relation}, one is
left with only a single independent form factor instead of three. We
choose $f_+(q^2)$ to be the single independent form factor. This is
helpful to formulate signature of NP in a manner where all hadronic
uncertainties are encoded into just the single form factor $f_+(q^2)$
for which the uncertainties are easily available from lattice QCD
estimates. Since in this work we do not primarily focus on the form
factors, we simply take both Eqs.~\eqref{eq:f0fp-relation} and
\eqref{eq:fpfT-relation} to be `empirically' true as depicted in
Fig.~\ref{fig:FFs-LQCD}, and the theoretical considerations that go
into analysing form factor relations can be found in
Ref.~\cite{KPRS:2024}. Before proceeding to the details of the new
physics signatures, we wish to point out that we have considered the
form factors predicted by HPQCD in \cite{Parrott:2022zte} which
suggest the simple relations Eqs.~\eqref{eq:f0fp-relation} and
\eqref{eq:fpfT-relation}. This allows us to do the full analysis using
a single form factor. One might as well consider the form factor fits
done in Ref.~\cite{Becirevic:2023aov}, which take into account
FNAL/MILC results as well as the HPQCD results for large $q^2$ region.

\section{New physics signatures in missing mass-square distribution of \texorpdfstring{$\boldsymbol{B \to K \, \nu \, \overline{\nu}}$}{B -> K nu nubar}}\label{sec:NP-signatures} 

Before delving into any further details on $B \to K \, \nu \,
\overline{\nu}$ we reiterate that we consider the final neutrinos to
be of the same flavor with non-zero mass\footnote{We note that it is
mathematically impossible to describe a massless chiral neutrino by a
real solution of Dirac equation and the real solution is essential for
Majorana neutrino, see Appendix B of \cite{Kim:2023iwz}. Nevertheless,
we emphasize that we \textit{neglect} all neutrino mass dependent
effects in our final  observables.} ($m_\nu \neq 0$) as already
experimentally observed. Due to th
e fact that, by definition, a
Majorana neutrino is quantum mechanically indistinguishable from its
corresponding antineutrino, the decay amplitude for such a case ought
to be anti-symmetric under the exchange of the two final neutrinos.
Additionally, due to the fact that both the neutrinos in the final
states can not be detected in the detector in the immediate vicinity
of the decay event, there are only two final observables in $B \to K
\, \nu \, \overline{\nu}$: the missing mass-square distribution and
the full partial branching ratio for the decay. In this section, we
consider both of these observables and study the signatures of new
physics in both of them.

\subsection{Decay amplitudes}

First considering the Dirac neutrino possibility, and assuming the
4-momentum of neutrino and the anti-neutrino to be $p_1$ and $p_2$
respectively, the decay amplitude for $B \to K \, \nu \,
\overline{\nu}$ is given by,
\begin{align}
&\mathscr{M}^D = \frac{4 \, G_F}{\sqrt{2}} \, \frac{\alphaem}{4\pi} \,
\lambda_t \nonumber\\%
&\times \Bigg( \left(p^\mu \, f_+ (q^2) - \left(\frac{m_B^2 -
m_K^2}{q^2}\right) \, q^\mu \, \left( f_+ (q^2) - f_0 (q^2) \right)
\right) \nonumber\\%
&\times \left( C_L^V \, \left[ \overline{u}(p_1) \, \gamma_\mu \, P_L
\, \varv (p_2)\right] + C_R^V \, \left[ \overline{u}(p_1) \,
\gamma_\mu \, P_R \, \varv (p_2)\right] \right) \nonumber\\%
&+ \left( \left(m_B + m_K\right) \, f_0 (q^2) \, C_L^S - \left(\frac{2
\, \big(p \cdot \left(p_1 - p_2\right)\big)}{m_B + m_K}\right) \,
f_T(q^2) \, C_L^T \right) \nonumber\\%
&\qquad \times \left[ \overline{u}(p_1) \, P_L \, \varv (p_2) \right]
\nonumber\\%
&+ \left( \left(m_B + m_K\right) \, f_0 (q^2) \, C_R^S - \left(\frac{2
\, \big(p \cdot \left(p_1 - p_2\right)\big)}{m_B + m_K}\right) \,
f_T(q^2) \, C_R^T \right) \nonumber\\%
&\qquad \times \left[ \overline{u}(p_1) \, P_R \, \varv (p_2) \right] 
\Bigg), \label{eq:Amp-D}
\end{align}
where we have made use of Gordon identities to get rid of the tensor
and axial-tensor contributions, and have neglected all those terms
which came out to be explicitly dependent on $m_\nu$.

For the case of Majorana neutrinos, as mentioned before
\cite{Kim:2022xjg}, the amplitude is anti-symmetric under exchange of
the two final neutrinos, and is given by,
\begin{align}
&\mathscr{M}^M = \frac{4 \, G_F}{2} \, \frac{\alphaem}{4\pi} \,
\lambda_t \nonumber\\%
&\times \Bigg( \left(p^\mu \, f_+ (q^2) - \left(\frac{m_B^2 -
m_K^2}{q^2}\right) \, q^\mu \, \left( f_+ (q^2) - f_0 (q^2) \right)
\right) \nonumber\\%
&\times \Big( \left(C_L^V - C_R^V\right) \, \left[ \overline{u}(p_1)
\, \gamma_\mu \, P_L \, \varv (p_2)\right]  \nonumber\\%
&\qquad + \left(C_R^V - C_L^V\right) \, \left[ \overline{u}(p_1) \,
\gamma_\mu \, P_R \, \varv (p_2)\right] \Big) \nonumber\\%
&+ 2 \, \left(m_B + m_K\right) \, f_0 (q^2) \, \Big( C_L^S \, \left[
\overline{u}(p_1) \, P_L \, \varv (p_2) \right] \nonumber\\%
&\qquad + C_R^S \, \left[ \overline{u}(p_1) \, P_R \, \varv (p_2)
\right] \Big) \Bigg), \label{eq:Amp-M}
\end{align}
where we have already introduced a factor $1/\sqrt{2}$ which takes
care of the statistical factor of $1/2$ at the stage of decay rate. In
Eq.~\eqref{eq:Amp-M}, the exchange properties of Majorana bilinears
have already been used for further simplification
\cite{Denner:1992vza}, which is also the reason for the absence of any
contributions from the tensor interactions in Majorana case.

\subsection{Missing mass-square distributions}

Taking the modulus square of the amplitudes of Eqs.~\eqref{eq:Amp-D}
and \eqref{eq:Amp-M}, neglecting all the terms that are explicitly
dependent on $m_\nu$ and doing the phase space integration, the
differential decay rate for $B \to K \, \nu \, \overline{\nu}$ in
terms of the missing mass-square $m_{\textrm{miss}}^2 \equiv q^2$ is
given by,
\begin{align}
\frac{\d \Gamma^{D/M}}{\d q^2} &= \frac{G_F^2\,\alphaem^2 \,
\modulus{\lambda_t}^2}{(2\,\pi)^5} \,
\frac{\sqrt{\lambda(q^2,m_B^2,m_K^2)}}{64 \, m_B^3} \nonumber\\%
&\quad \times \left( T_0^{D/M} + \frac{1}{3} \, T_2^{D/M} \right) \,
\sqrt{1-\frac{4\,m_\nu^2}{q^2}}, \label{eq:m2miss_distribution}
\end{align}
where the various terms in the differential decay rate are,
\begin{subequations}\label{eq:DM-terms}
\begin{align}
T_0^D &= q^2 \, \left(m_B+m_K\right)^2 \, \left(\left(C^S_L\right)^2 +
\left(C^S_R\right)^2\right) \, f_0^2(q^2) \nonumber\\%
&\quad + \lambda(q^2,m_B^2,m_K^2) \, \left(\left(C^V_L\right)^2 +
\left(C^V_R\right)^2\right) \, f_+^2(q^2), \label{eq:Dirac-term-0}\\%
T_2^D &= \lambda(q^2,m_B^2,m_K^2) \,
\Bigg(\frac{4\,q^2\,\left(\left(C^T_L\right)^2 +
\left(C^T_R\right)^2\right) \, f_T^2(q^2)}{\left(m_B + m_K\right)^2}
\nonumber\\%
&\quad - \left(\left(C^V_L\right)^2 + \left(C^V_R\right)^2\right) \,
f_+^2(q^2) \Bigg), \label{eq:Dirac-term-2}\\%
T_0^M &= 2 \, q^2 \,
\left(m_B+m_K\right)^2\,\left(\left(C^S_L\right)^2 +
\left(C^S_R\right)^2\right) \, f_0^2(q^2) \nonumber\\%
&\quad + \lambda(q^2,m_B^2,m_K^2) \, \left(C^V_L - C^V_R\right)^2 \,
f_+^2(q^2), \label{eq:Majorana-term-0}\\%
T_2^M &= - \lambda(q^2,m_B^2,m_K^2) \, \left(C^V_L - C^V_R\right)^2 \,
f_+^2(q^2), \label{eq:Majorana-term-2}%
\end{align}
\end{subequations}
with the K\"{a}ll\'{e}n function
$\lambda(x,y,z)$ being given by,
\begin{equation}
\lambda(x,y,z) = x^2 + y^2 + z^2 - 2\,\big(x\,y+y\,z+z\,x\big).
\end{equation}
From Eq.~\eqref{eq:DM-terms} it is easy to infer that the presence of
right-chiral vector NP and scalar NP can, in principle, contribute
differently to Dirac and Majorana scenarios. As noted before, the
tensor current interactions do not affect the Majorana case
distribution.

\subsubsection{Missing mass-square distribution in the SM}

\begin{figure}[hbtp]
\centering%
\includegraphics[scale=0.8]{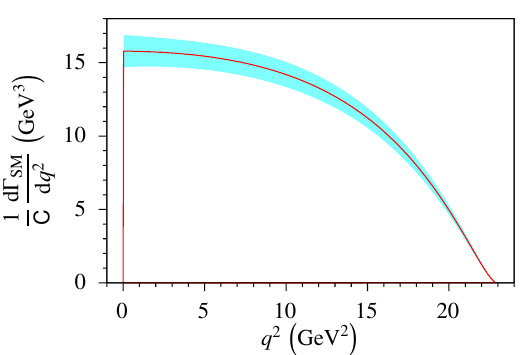}%
\caption{The trend in the missing mass-square ($q^2$) distribution in
the SM. The cyan colored band around the central red line corresponds
to $1$ standard deviation in the form factor $f_+(q^2)$ taken from the
prediction by the HPQCD collaboration \cite{Parrott:2022zte}.}%
\label{fig:SM-q2-trend}
\end{figure}

\textit{In the SM}, we satisfy Eq.~\eqref{eq:SM-Wilson-coeff}, which
upon substitution into Eq.~\eqref{eq:m2miss_distribution} clearly
shows \textit{no observable difference between Dirac and Majorana
neutrinos in $B^+ \to K^+ \, \nu \, \overline{\nu}$ decay}. Therefore,
in the SM we have,
\begin{align}
\frac{\d\Gamma_\textrm{SM}}{\d q^2} &\equiv
\frac{\d\Gamma^{D}_\textrm{SM}}{\d q^2} =
\frac{\d\Gamma^{M}_\textrm{SM}}{\d q^2} \nonumber\\%
&= \frac{G_F^2\,\alphaem^2 \, \modulus{\lambda_t}^2}{(2\,\pi)^5} \,
\frac{\sqrt{\lambda(q^2,m_B^2,m_K^2)}}{64 \, m_B^3} \nonumber\\%
&\quad \times \left( \left(T_0\right)_{\textrm{SM}} + \frac{1}{3} \,
\left(T_2\right)_{\textrm{SM}} \right) \,
\sqrt{1-\frac{4\,m_\nu^2}{q^2}} \, ,
\label{eq:m2miss-distribution-SM-1}
\end{align}%
where
\begin{subequations}
\begin{align}
\left(T_0\right)_{\textrm{SM}} &= \lambda(q^2,m_B^2,m_K^2) \,
\left(C^V_L\right)_{\textrm{SM}}^2 \, f_+^2(q^2) \, ,\\%
\left(T_2\right)_{\textrm{SM}} &= - \lambda(q^2,m_B^2,m_K^2) \,
\left(C^V_L\right)_{\textrm{SM}}^2 \, f_+^2(q^2) \equiv -
\left(T_0\right)_{\textrm{SM}}.
\end{align}
\end{subequations}
Thus, the SM differential decay rate, or the missing mass-square
distribution, is dependent on the vector form factor $f_+(q^2)$ alone,
\begin{equation}\label{eq:m2miss-distribution-SM-2}
\frac{\d\Gamma_\textrm{SM}}{\d q^2} = \mathsf{C} \,
\frac{\left(\lambda(q^2,m_B^2,m_K^2)\right)^{3/2}}{m_B^3}  \,
\sqrt{1-\frac{4\,m_\nu^2}{q^2}} \, f_+^2(q^2),
\end{equation}
where $\mathsf{C}$ is the constant factor given by,
\begin{equation}
\mathsf{C} = \frac{G_F^2\,\alphaem^2 \,
\modulus{\lambda_t}^2}{(2\,\pi)^5} \,
\frac{\left(C^V_L\right)_{\textrm{SM}}^2}{96}.
\end{equation}
The $q^2$ dependence in the SM can be seen from
Fig.~\ref{fig:SM-q2-trend} which plots $\dfrac{1}{\mathsf{C}} \,
\dfrac{\d\Gamma_\textrm{SM}}{\d q^2}$, where the cyan band corresponds
to the $1\sigma$ error in $f_+(q^2)$ as predicted by the HPQCD
collaboration~\cite{Parrott:2022zte}.

\subsubsection{Missing mass-square distribution for generic NP possibilities with only one form factor}

The Eq.~\eqref{eq:m2miss_distribution} is the most general expression
for the missing mass-square distribution for generic NP possibilities
(i.e.\ it does not consider a specific UV complete NP model, and thus
is model-independent from this perspective). However, as we have noted
before we can relate the three form factors using
Eqs.~\eqref{eq:f0fp-relation} and \eqref{eq:fpfT-relation} and express
the general missing mass-square distribution in terms of a single form
factor, which we choose to be $f_+(q^2)$, the same form factor which
appears in the SM, see Eq.~\eqref{eq:m2miss-distribution-SM-2}. Doing
this, the expressions for $T_{0,2}^D$ and $T_0^M$ get modified to the
following form,
\begin{subequations}\label{eq:NP-DM-terms}
\begin{align}
T_0^D &= \Bigg[ \left(\left(C^S_L\right)^2 +
\left(C^S_R\right)^2\right) \, \frac{q^2 \, \left( (m_B + m_K)^2 - q^2
\right)^2}{(m_B + m_K)^2} \nonumber\\%
&\quad + \left(\left(C^V_L\right)^2 + \left(C^V_R\right)^2\right) \,
\lambda(q^2,m_B^2,m_K^2) \Bigg] \, f_+^2(q^2),
\label{eq:NP-Dirac-term-0}\\%
T_2^D &= \lambda(q^2,m_B^2,m_K^2) \,
\Bigg(\frac{4\,q^2\,\left(\left(C^T_L\right)^2 +
\left(C^T_R\right)^2\right)}{\left(m_B + m_K\right)^2} \nonumber\\%
&\quad - \left(\left(C^V_L\right)^2 + \left(C^V_R\right)^2\right)
\Bigg) \, f_+^2(q^2), \label{eq:NP-Dirac-term-2}\\%
T_0^M &= \Bigg[ \left(\left(C^S_L\right)^2 +
\left(C^S_R\right)^2\right) \, \frac{2 \,q^2 \, \left( (m_B+m_K)^2 -
q^2 \right)^2}{(m_B+m_K)^2} \nonumber\\%
&\quad + \left(C^V_L - C^V_R\right)^2 \, \lambda(q^2,m_B^2,m_K^2)
\Bigg] \, f_+^2(q^2), \label{eq:NP-Majorana-term-0}
\end{align}
\end{subequations}
The expression for $T_2^M$ is already proportional to $f_+^2(q^2)$ in
Eq.~\eqref{eq:Majorana-term-2}, so it remains unchanged. 

Since the missing mass-square distribution for both the generic NP
case and the SM case are both dependent on the same single form factor
$f_+(q^2)$, we can parameterize the effects of NP in a manner which
does not depend on the hadronic uncertainties inherent in the form
factors. Before doing that we introduce the following ``effective
$\varepsilon$ parametrization'' of NP (similar to the approach in
\cite{Kim:2022xjg}) which are agnostic to Dirac or Majorana
nature of neutrino and only parameterize the small deviation from SM
Wilson coefficient in terms of $\left( C_L^V \right)_\text{SM}$, as
explained before,
\begin{subequations}\label{eq:eps-parameters}
\begin{align}
C^S_{L} &= \varepsilon^S_{L} \, \left(C^V_L\right)_{\textrm{SM}},\\%
C^S_R &= \varepsilon^S_R\, \left(C^V_L\right)_{\textrm{SM}},\\%
C^V_L &= \left( 1+ \varepsilon^V_L\right) \,
\left(C^V_L\right)_{\textrm{SM}},\\%
C^V_R &= \varepsilon^V_R \, \left(C^V_L\right)_{\textrm{SM}},\\%
C^T_{L} &= \varepsilon^T_{L} \, \left(C^V_L\right)_{\textrm{SM}},\\%
C^T_R &= \varepsilon^T_R \, \left(C^V_L\right)_{\textrm{SM}},%
\end{align}
\end{subequations}
where all the $\varepsilon_X^Y$'s (for $X=L,R$ and $Y=S,V,T$) are real
but arbitrary constants (i.e.\ not dependent on $q^2$ in any manner
and arbitrary in absence of a specific UV complete NP model as done in
our minimally extended model-independent approach). In terms of these
$\varepsilon$ parameters we can rewrite the generic missing
mass-square distribution as follows,
\begin{subequations}\label{eq:R}
\begin{align}
\dfrac{\d \Gamma^D}{\d q^2} &= \dfrac{\d\Gamma_\textrm{SM}}{\d q^2} \;
\Bigg( R^D_S(q^2) \, \left(\left(\varepsilon^S_L\right)^2 +
\left(\varepsilon^S_R\right)^2 \right) \nonumber\\*%
&\quad + R^D_V(q^2) \, \left( \left(1+\varepsilon^V_L\right)^2 +
\left(\varepsilon^V_R\right)^2 \right) \nonumber\\*%
&\quad + R^D_T(q^2) \, \left( \left(\varepsilon^T_L\right)^2 +
\left(\varepsilon^T_R\right)^2 \right) \Bigg), \label{eq:RD}\\%
\dfrac{\d \Gamma^M}{\d q^2} &= \dfrac{\d\Gamma_\textrm{SM}}{\d q^2} \;
\Bigg( R^M_S(q^2) \, \left(\left(\varepsilon^S_L\right)^2 +
\left(\varepsilon^S_R\right)^2 \right) \nonumber\\%
&\quad + R^D_V(q^2) \, \left( \left(1+\varepsilon^V_L -
\varepsilon^V_R\right)^2 \right) \Bigg), \label{eq:RM}%
\end{align}
\end{subequations}
where the ``NP weight factors'' $R^{D/M}_Y(q^2)$ are
\textit{independent of form factor uncertainties} (all relevant
hadronic uncertainty is present only in $f_+^2(q^2)$ which appears in
the SM differential decay rate, as shown in
Eq.~\eqref{eq:m2miss-distribution-SM-2}), and these are given by,
\begin{subequations}\label{eq:NP-weights}
    \begin{align}
        R^D_S(q^2) &= \frac{3 \, q^2 \, \left((m_B + m_K)^2 -
        q^2\right)^2}{2 \, \lambda(q^2,m_B^2,m_K^2) \, (m_B + m_K)^2}
        = \frac{1}{2} R^M_S(q^2),\label{eq:NP-weights-SDM}\\%
        R^D_V(q^2) &= 1 = R^M_V(q^2), \label{eq:NP-weights-VDM}\\%
        R^D_T(q^2) &= \frac{2\,q^2}{(m_B + m_K)^2},
        \label{eq:NP-weights-TD}\\%
        R^M_T(q^2) &= 0. \label{eq:NP-weights-TM}
    \end{align}
\end{subequations}

\begin{figure}[hbtp]
\centering%
\includegraphics[width=\linewidth,keepaspectratio]{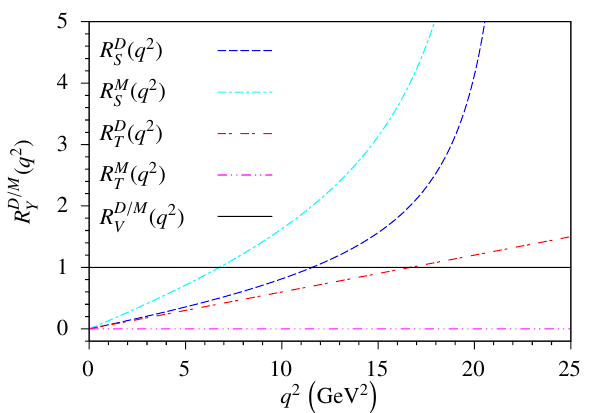}%
\caption{Comparison of $q^2$ dependence of scalar, vector and tensor
NP weight factors $R_Y^{D/M}(q^2)$ for $Y=S,V,T$, which appear in
Eqs.~\eqref{eq:RD} and \eqref{eq:RM}, and as given in
\eqref{eq:NP-weights}. Note that $R_T^M(q^2) = 0$ is simply due
to the fact that there is no allowed tensor interaction for our
Majorana case.} \label{fig:NP-weights}
\end{figure}

Note that the contribution of scalar NP has similar $q^2$ dependence
in both Dirac and Majorana neutrino possibilities, but the effects are
two times stronger in Majorana case than the Dirac case, see
Eq.~\eqref{eq:NP-weights-SDM} and Fig.~\ref{fig:NP-weights}. The
scalar NP weight factor has a divergence\footnote{This divergence is
not present in the final differential decay rate, since the factor
$\lambda(q^2,m_B^2,m_K^2)$ in denominator of
Eq.~\eqref{eq:NP-weights-SDM} gets countered by the factor
$\left(\lambda(q^2,m_B^2,m_K^2)\right)^{3/2}$ in the numerator of
Eq.~\eqref{eq:m2miss-distribution-SM-2} which ensures that the full
differential decay rate is analytic in the allowed $q^2$ range.} at
the kinematic end point as $\lambda(q^2,m_B^2,m_K^2)$ present in
denominator of Eq.~\eqref{eq:NP-weights-SDM} vanishes at
$q^2=(m_B-m_K)^2$. The tensor NP contribution only affects the Dirac
case, and the effect increases linearly with $q^2$, see
Fig.~\ref{fig:NP-weights}. The different $q^2$ dependence of the
various NP possibilities, suggests that one can distinguish them from
one another if one can experimentally study the $q^2$ distribution.
This has been highlighted in Ref.~\cite{Kim:2022xjg}, where for
simplicity, the $q^2$ dependencies of the form factors were not taken
into consideration. However, our expressions Eqs.~\eqref{eq:R} and
\eqref{eq:NP-weights} take into account the form factor relations of
Eqs.~\eqref{eq:f0fp-relation} and \eqref{eq:fpfT-relation} which fit
the lattice QCD form factor predictions very well, as shown in
Fig.~\ref{fig:FFs-LQCD}. We expect that after the recent observation
of the rare decay $B \to K \, \nu \, \overline{\nu}$, the experimental
study of $q^2$ distribution in this decay would also be available in
future. This would be crucial to study new physics in a more precise
manner.

It is also easy to see in Fig.~\ref{fig:NP-weights} that the NP weight
factors $R_Y^{D/M}(q^2)$ (with $Y=S,T$) become greater than the SM
weight factor $R_V^{D/M} = 1$ in some regions of $q^2$. As mentioned
before, the $\varepsilon$ parameters are constants which are not
dependent on $q^2$. Therefore, in context of our SM allowed decay, if
one is interested in NP contributions to be smaller than the SM
contribution for any value of $q^2$, then the $\varepsilon$ parameters
must, in general, satisfy
\begin{equation}\label{eq:physical-limit-epsilons}
-1 < \varepsilon_X^Y < 1.
\end{equation}
Nonetheless, while comparing experimental measurement or data with the
SM prediction to obtain an estimate of NP $\varepsilon$'s, one can
relax Eq.~\eqref{eq:physical-limit-epsilons}. For clarity and
convenience, we will use the terminology of ``\textit{inner region}''
for $\varepsilon$'s satisfying Eq.~\eqref{eq:physical-limit-epsilons}
and ``\textit{outer region}'' for values of $\varepsilon$'s beyond it.
We reiterate that the $\varepsilon$ parameters as defined in
Eq.~\eqref{eq:eps-parameters} are agnostic to the Dirac or Majorana
natures of the neutrinos. Hence there are no \textit{a priori}
correlations over the $\varepsilon$'s when comparing Dirac and
Majorana neutrino possibilities.

We note that, if one considers only vector NP contributions alone,
then from Eqs.~\eqref{eq:RD} and \eqref{eq:RM}, it is straightforward
to show that
\begin{equation}\label{eq:D-M-NPV}
    \dfrac{\d \Gamma^D}{\d q^2} - \dfrac{\d \Gamma^M}{\d q^2} =
    \dfrac{\d\Gamma_\textrm{SM}}{\d q^2} \; \left(2\varepsilon_R^V \,
    \left(1+\varepsilon_L^V\right) \right), \;
    \left(\text{\parbox{2cm}{\centering considering only vector NP
    possibility}}\right)
\end{equation}
which implies that for $\varepsilon_R^V >0$ we get $\dfrac{\d
\Gamma^D}{\d q^2} > \dfrac{\d \Gamma^M}{\d q^2}$ in the whole inner region for $\varepsilon_L^V$, i.e.\ $-1 < \varepsilon_L^V \leqslant 1$.
From Eq.~\eqref{eq:D-M-NPV} it is also clear that left-handed vector
NP contribution alone would not lead to any difference between the
Dirac and Majorana possibilities, one needs non-zero right-handed
vector NP contribution for this purpose. In fact, the difference
between Dirac and Majorana cases would also exist even if one
considers only right-handed vector NP possibility, i.e.\
$\varepsilon_R^V \neq 0$ and $\varepsilon_L^V=0$.

\begin{figure*}[ht!]
\centering%
\includegraphics[scale=0.8]{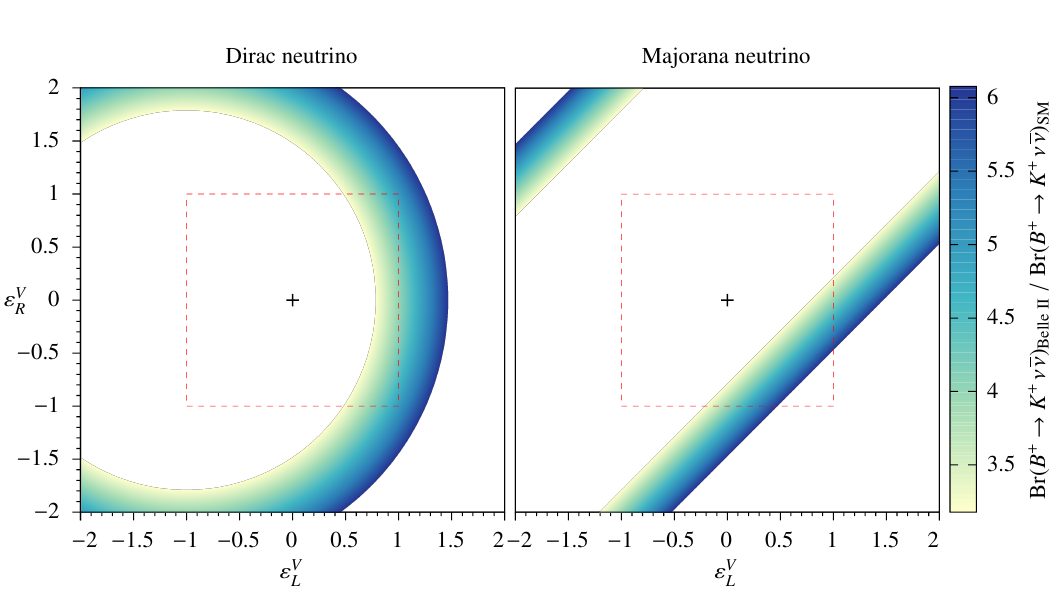}%
\caption{The parameter space of $\varepsilon_L^V$ and
$\varepsilon_R^V$ constrained by the Belle-II measurement of the
branching ratio of $B^+ \to K^+ \, \nu \, \overline{\nu}$. The SM
point corresponding to $\varepsilon_L^V = 0 = \varepsilon_R^V$ is
shown by the `+' symbol. The shaded regions correspond to the
$1\sigma$ band for the ratio $\textrm{Br}(B^+ \to K^+ \, \nu \,
\overline{\nu})_{\textrm{Belle~II}}/\textrm{Br}(B^+ \to K^+ \, \nu \,
\overline{\nu})_{\textrm{SM}}$, with errors propagated by quadrature.
Only for Majorana case the entire $1\sigma$ region can be found in the
inner region of the parameter space which is the interior of the red
dashed square.}%
\label{fig:Constraints-VDM}
\end{figure*}

\begin{figure*}[ht!]
\centering%
\subfloat[Constraint on scalar NP contribution
\label{fig:scalar-NP}]{\includegraphics[scale=0.8]{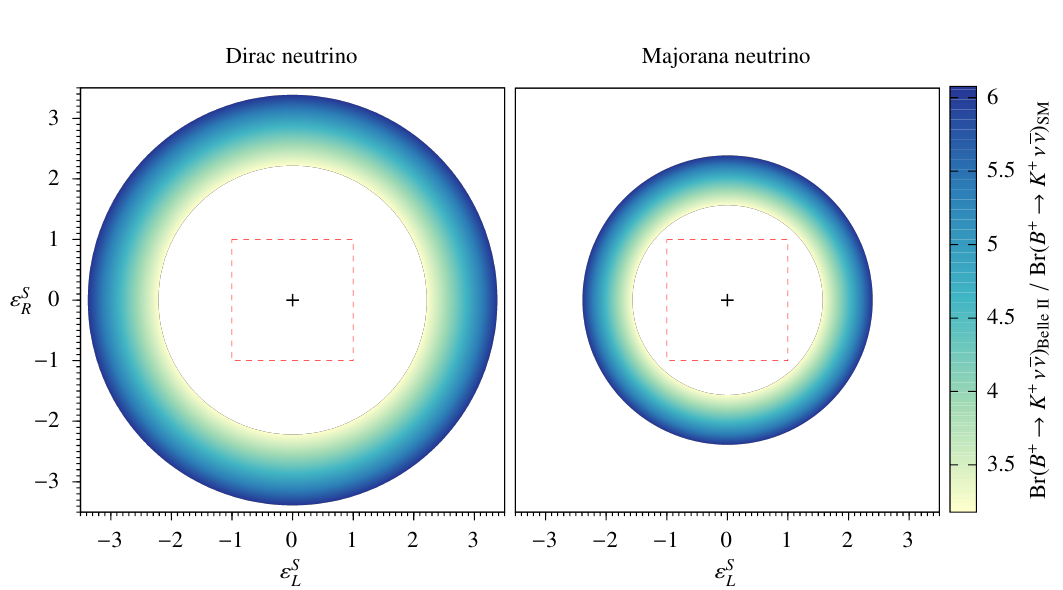}}\\%
\subfloat[Constraint on tensor NP contribution
\label{fig:tensor-NP}]{\includegraphics[scale=0.8]{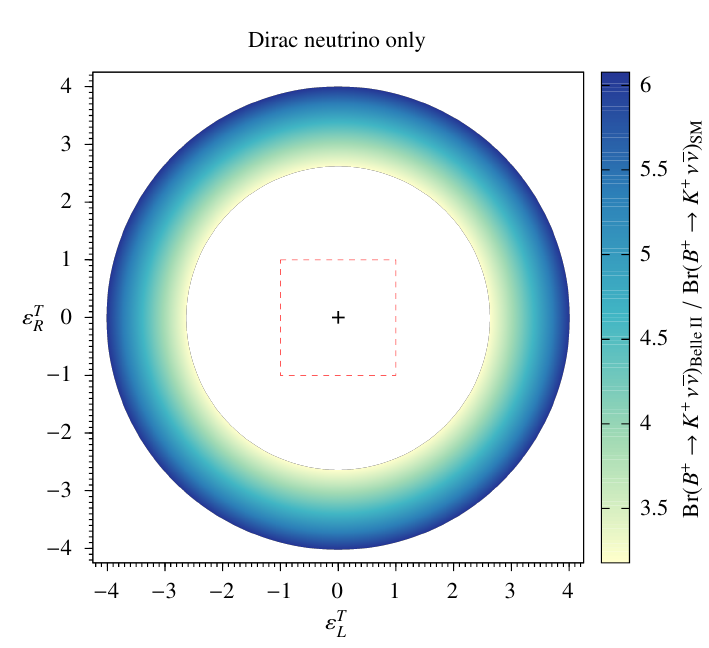}}%
\caption{The parameter space of $\varepsilon_L^S$ and
$\varepsilon_R^S$, as well as $\varepsilon_L^T$ and $\varepsilon_R^T$,
constrained by the Belle~II measurement of the branching ratio of
$B^+\to K^+ \, \nu \, \overline{\nu}$. The SM point corresponding to
$\varepsilon_L^V = 0 = \varepsilon_R^V$ is shown by the `+' symbol.
The shaded regions correspond to the $1\sigma$ band for the ratio
$\textrm{Br}(B^+ \to K^+ \, \nu \,
\overline{\nu})_{\textrm{Belle~II}}/\textrm{Br}(B^+ \to K^+ \, \nu \,
\overline{\nu})_{\textrm{SM}}$, with errors propagated by quadrature.
No tensor NP contribution is possible for Majorana neutrinos. The
interior of the red dashed square shows the inner region of the
parameter space for which NP contribution is expected to be smaller
than the SM contribution.}%
\label{fig:scalar-tensor-NP}
\end{figure*}

As shown in Fig.~\ref{fig:NP-weights} and as is clear from
Eqs.~\eqref{eq:R} and \eqref{eq:NP-weights-VDM}, the vector NP
contributions do not yield any further $q^2$ dependence than what is
already present in the SM. Therefore, if we were to \textit{assume}
that in addition to SM \textit{only vector NP contributes} to
$B \to K \, \nu_\ell \, \overline{\nu}_\ell$ (due to whatever reasons
one might have, e.g.\ in a model with $Z'$), it is straightforward to
note that,
\begin{align}\label{eq:BrNPvsSM}
&  \dfrac{\Br\left( B \to K \, \nu_\ell \, \overline{\nu}_\ell
\right)_{\textrm{NP}+\textrm{SM}}}{\Br\left( B \to K \, \nu_\ell \,
\overline{\nu}_\ell \right)_{\textrm{SM}}} = 
\begin{cases}
\left(1+\varepsilon^V_L\right)^2 + \left(\varepsilon^V_R\right)^2, &
\left( \parbox{15mm}{\centering for Dirac neutrinos} \right) \\[5mm]%
\left(1+\varepsilon^V_L-\varepsilon^V_R\right)^2. &  \left(
\parbox{15mm}{\centering for Majorana neutrinos} \right) %
\end{cases}
\end{align}

Interestingly, as we noted in introduction in Eqs.~\eqref{eq:BrSM} and
\eqref{eq:BrBelleII}, the Belle-II measurement of the branching ratio
of $B^+ \to K^+ \, \nu \, \overline{\nu}$ is different from the SM
prediction. Therefore, if we consider
\begin{equation}
\Br\left( B \to K \, \nu_\ell \, \overline{\nu}_\ell
\right)_{\textrm{NP}+\textrm{SM}} = \Br\left( B \to K \, \nu_\ell \,
\overline{\nu}_\ell \right)_{\textrm{Belle~II}},
\end{equation}
and use Eq.~\eqref{eq:BrNPvsSM} to estimate the possible contributions
from vector NP possibility alone which can accommodate the observed
discrepancy. The parameter space of $\varepsilon^V_L$ and
$\varepsilon^V_R$ is scanned in the inner ($-1 \leqslant
\varepsilon^V_{L,R} \leqslant 1$) and outer ($-1 >
\varepsilon^V_{L,R}$ or $\varepsilon^V_{L,R} > 1$) regions in
Fig.~\ref{fig:Constraints-VDM}. As shown by the range of the
color-bars, the constrained (shaded) regions correspond to the
$1\sigma$ range of the branching ratio as measured by the Belle-II
collaboration. Clearly, for the Dirac neutrino hypothesis, no portion
of the full $1\sigma$ band of $\Br\left( B \to K \, \nu_\ell \,
\overline{\nu}_\ell \right)$ is completely inside inner region, while
for Majorana neutrino hypothesis we do find a part of the allowed
$1\sigma$ band to be fully inside the inner region, i.e. $-0.2 <
\varepsilon^V_{L} < 1, ~ -1 < \varepsilon^V_{R} < 0.2$. At the current
level of experimental precision, it might nevertheless be premature to
conclude that the Majorana nature is more preferred than the Dirac
nature. However, this would be an interesting feature to study in
future, when a new improved measurement of the branching ratio of $B^+
\to K^+ \, \nu \, \overline{\nu}$ gets reported.

If instead of vector NP contribution, we assume that the scalar or
tensor NP contributions alone can account for the Belle-II observed
branching ratio in Eq.~\eqref{eq:BrBelleII}, then considering either
scalar or tensor NP contributions one at a time, we find that,
\begin{subequations}
\begin{align}
\left(\varepsilon^S_L\right)^2 + \left(\varepsilon^S_R\right)^2 &= 
\begin{cases}
8.14 \pm 3.25, & \textrm{(for Dirac neutrinos)}\\%
4.07 \pm 1.63, & \textrm{(for Majorana neutrinos)}
\end{cases}\\%
\left(\varepsilon^T_L\right)^2 + \left(\varepsilon^T_R\right)^2 &=
11.45 \pm 4.58 \quad \textrm{(only for Dirac neutrinos)}.
\end{align}
\end{subequations}
The allowed regions inside the parameter space of $\varepsilon_L^S$
and $\varepsilon_R^S$, as well as $\varepsilon_L^T$ and
$\varepsilon_R^T$, are found to be fully in the outer region (i.e. $-1
\leqslant \varepsilon^{S,T}_{L,R} \leqslant 1$), see
Fig.~\ref{fig:scalar-tensor-NP}. These suggest that one would require
that the contributions from either scalar or tensor type NP
interactions be necessarily larger than the SM contribution to account
for the present branching ratio reported by the Belle~II
collaboration. It would be exciting for specific model-dependent NP
searches, if any future measurements of the branching ratio of $B^+
\to K^+ \, \nu \, \overline{\nu}$ show preference for scalar and/or
tensor NP contributions. Rather than the branching ratio measurements,
it is the missing mass-square ($q^2$) distribution which would be able
to distinguish among the scalar, vector and tensor NP possibilities,
as they have different $q^2$ dependencies. Before we conclude, we
would like to reiterate that our model-independent analysis and the
results discussed above are strictly in context of both lepton flavor
conservation and lepton flavor universality.

\section{Conclusion}\label{sec:conclusion}

We have studied the prospects of generically probing NP in the $B^+
\to K^+ \, \nu \, \overline{\nu}$ decay in light of the recent
measurement of the branching ratio of this process by the Belle-II
collaboration. We consider NP interactions involving generic scalar,
vector and tensor type of interactions. The scalar form factor
$f_0(q^2)$, vector form factor $f_+(q^2)$ and tensor form factor
$f_T(q^2)$ play very significant roles in calculating the decay width,
since the hadronic uncertainties are majorly contributed by these,
although other intermediate resonance effects might also be
non-avoidable such as the effect of $J/\psi$ in case of $B \to K \,
\ell^+ \, \ell^-$. We have utilized a set of simple analytical
expressions which relate the $B \to K$ form factors and these
relations are highly in agreement with the predictions from the fully
relativistic lattice QCD calculations reported by the HPQCD
collaboration.

Using these form factor relations to our advantage we proceed to do a
careful study of generic NP possibilities in the decay $B \to K \, \nu
\, \overline{\nu}$ in which no intermediate resonances such as
$J/\psi$ play any adverse role. We study the NP signatures for both
the Dirac and Majorana neutrino hypotheses. Interestingly, if only
vector NP contribution is assumed to be contributing to $B^+ \to K^+
\, \nu \, \overline{\nu}$ (due to whatever fundamental reasons one
might have in one's favorite UV complete NP model, such as say a model
with a $Z'$), then the recently reported measurement of branching
ratio of $B^+ \to K^+ \, \nu \, \overline{\nu}$ by the Belle-II
collaboration shows that the Majorana neutrino  can fully account for
the entire $1\sigma$ range of the experimental observation within the
`inner region', a generic domain of the NP parameters for which SM
contribution is larger than the NP contribution. The Dirac neutrino
hypothesis does not fully account for the observed branching ratio in
the this inner region of parameter space. However, it remains to be
seen if in future more precise branching ratio measurements would ever
rule out the Dirac neutrino hypothesis. If one assumes the NP
interactions to be either scalar or tensor type in nature, very large
values of NP parameters are required to reconcile with the Belle-II
measurement, for both Dirac and Majorana cases. More precise future
measurements of branching ratio as well as missing-mass square
distribution of $B \to K \, \nu \, \overline{\nu}$ decay would help in
exploration of various new physics possibilities in the context of
Dirac or Majorana nature of the neutrinos.

\section*{Acknowledgements}

The work of CSK is supported through the National Research Foundation
of Korea (NRF-2022R1I1A1A01055643, \newline  NRF-2022R1A5A1030700).
The work of DS is supported by the Polish National Science Centre
under the Grant number DEC-2019/35/B/ST2/02008. V.K.N. is supported by
ANID-Chile Fondecyt Postdoctoral grant 3220005. DS is grateful to
Professors Stefan Pokorski and Janusz Rosiek for discussions on the
form factors.

\bibliographystyle{utphys_jr} 
\bibliography{References}

\end{document}